\documentclass[amsmath,amssymb,twocolumn]{revtex4}
\usepackage{subfig}
\usepackage{dcolumn}
\usepackage{bm}
\usepackage{amsmath,mathrsfs}
\usepackage{graphicx,epsfig}
\usepackage{hyperref}
\usepackage{epsf}
\usepackage{color}

\begin{document}

%-----------------------------------------------------------------------

\title{{\bf Non-exotic asymptotically flat wormholes in $f(Q,T)$ gravity}}

\author{S. Rastgoo}\email{rastgoo@sirjantech.ac.ir}
\author{ F. Parsaei }\email{fparsaei@gmail.com}
\author{S. Nasirimoghadam}\email{soudabe.nasirimoghadam@gmail.com}

\affiliation{ Physics Department , Sirjan University of Technology, Sirjan 78137, Iran.}

\date{\today}

%-----------------------------------------------------------------------

\begin{abstract}
\par   In this study, we investigate the possible existence of static and spherically symmetric wormhole solutions within the context of the newly formulated extended $f(Q,T)$ gravity. We analyze a linear model, $f(Q,T)=\alpha Q+ \beta T$, and focus on traversable wormholes. By applying the variational method, we derive modified versions of the field equations that are influenced by an anisotropic matter source for a zero redshift function. It has been observed that the violation of energy conditions is influenced by the parameters $\alpha$ and $\beta$.  We reach the conclusion that solutions which violate the radial and lateral null energy condition in the context of general relativity may still adhere to the energy conditions within the realm of $f(Q,T)$ gravity. To begin with, by utilizing a linear equation of state for radial pressure, we obtain a power-law shape function. Additionally, we investigate solutions defined by a variable equation of state parameter. A broad spectrum of non-exotic wormhole solutions has been identified, contingent upon the particular parameters of the model. \\
 %\textbf{Keywords} : wormholes, f(Q), energy conditions
\end{abstract}

\maketitle
\section{Introduction}

The term 'wormhole' was first introduced in \cite{wheeler} by Misner and Wheeler. Although the concept is fascinating, these structures remain entirely theoretical and necessitate further theoretical exploration. Later, Morris and Thorne \cite{WH} reinterpreted this concept by integrating traversable wormhole solutions into the Einstein field equations, allowing for a safe passage through the wormhole without the hindrance of event horizons.
Within the context of general relativity (GR), the presence of exotic matter is essential for maintaining the stability of a wormhole's throat \cite{Visser}. Energy conditions (ECs) play a vital role in GR as they impose limitations on the types of matter and energy permissible in the universe. They assist physicists in comprehending the stability and causality of spacetime geometries. The breach of the null energy condition (NEC) is a fundamental aspect of wormhole theory within the framework of GR. To this day, GR remains the prevailing theory supported by many due to its significant impact on gravitational forces, as demonstrated by numerous observations and experiments. Nevertheless, there are particular phenomena governed by GR that it fails to adequately explain, such as the accelerated expansion of the universe, gravitational interactions at galactic scales, and the ongoing pursuit of a quantum theory that can precisely characterize gravity.
In this context, scholars have presented a range of concepts that can be classified into two separate categories of altered propositions: those related to altered matter and those linked to altered curvature. Phantom fluid, acknowledged as a specific form of dark energy (DE), can be considered a type of exotic matter.
This idea was first proposed to explain the accelerated expansion of the universe, which was subsequently followed by extensive research on phantom wormholes that contravene ECs, as noted in the existing literature \cite{phantom, phantom2, phantom1}. Considering that exotic matter is deemed unlikely within the framework of GR, numerous researchers have endeavored to reduce its necessity. These strategies aim to diminish the reliance on exotic matter while still satisfying the conditions required for spacetime geometry. In these theoretical frameworks, exotic matter is confined to specific areas of spacetime. Prominent examples include thin-shell wormholes \cite{cut, cut2, cut3}, those characterized by a variable equation of state (EoS) \cite{Remo, variable}, and wormholes defined by a polynomial EoS \cite{foad}.

As the alternative approach, modifications to the traditional theory of gravity, particularly GR, have mainly been motivated by the limitations of GR in accounting for the accelerated expansion of the universe during later epochs. The impact of modifying GR on wormhole theory can be profound, as such changes can introduce new dynamics and properties to the solutions of Einstein's equations. In some modified theories, the constraints imposed by traditional ECs may be altered, allowing for the existence of wormholes that do not necessitate exotic matter. This can lead to the discovery of new types of wormholes, which could be more physically plausible. Additionally, these modifications may change the stability criteria for wormholes, potentially allowing for configurations that are stable under certain conditions, thus enhancing their theoretical viability within the framework of modern physics. Wormholes are explored in various theoretical frameworks, including Braneworld \cite{b1, b2, b3}, Born-Infeld theory \cite{Bo, Bo1}, quadratic gravity \cite{quad, quad1}, Einstein-Cartan gravity \cite{Cartan, Cartan1, Cartan2}, Rastall gravity \cite{Rast}, Rastall–Rainbow gravity \cite{RaR, RaR1},  $f(Q)$ gravity \cite{fq, fq2, fq4, fq44, fq5}, $f(R)$ gravity \cite{Nojiri, fR11, fR22, fR55}, Ricci inverse gravity \cite{inverse}, $f(T,\mathcal{T})$ gravity \cite{Must1, Err, Riz, Must2, foad3}, $f(R, l_m)$ gravity \cite{L1, L2, L7, L10, L12, SR2} and $f(R,T)$ gravity \cite{Azizi, Moa, Zub, fr2, Sha, Ban, Sarkar, SR1, foad4}. As a result of the alteration of Einstein's field equations in these contexts, certain scenarios are able to address the issue of exotic matter within the framework of wormhole theory.

Lagrangian-based theories of gravity use the principles of Lagrangian mechanics to formulate the dynamics of gravitational fields. The fundamental idea is to construct a Lagrangian density that encapsulates the gravitational interactions and then derive the equations of motion through the principle of least action. In Einstein's GR the Lagrangian density is given by the Ricci scalar $R$, which is derived from the geometry of spacetime. The Einstein-Hilbert action is expressed as: $ S=\int{R\sqrt{-g}\text{d}^4x}$,
where $g$ is the determinant of the metric tensor. The equations of motion are obtained by varying this action with respect to the metric tensor, leading to Einstein's field equations. This formulation highlights the geometric nature of gravity and its relation to the curvature of spacetime. Many alternative theories of gravity, such as $f(R)$ gravity \cite{Ha}, $f(\mathcal{T})$ gravity \cite{f(T)}, and others, extend the Lagrangian framework by introducing functions of the Ricci scalar ($R$) or torsion ($\mathcal{T}$).  This modification allows for additional degrees of freedom and can lead to different cosmological and gravitational phenomena, such as late-time cosmic acceleration without invoking DE.
In addition to previous extensions of GR, we highlight the $f(Q)$ gravitational theory introduced by Jimenez et. al. \cite{f(Q)}. In this theory, both torsion and curvature will vanish, and thus gravity is solely dependent on the non-metricity scalar $Q$. Yixin et al. \cite{fqt} introduced an extension of $f(Q)$ gravity referred to as $f(Q,T)$ gravity, which is founded on the interaction between the non-metricity $Q$ and the trace of the energy-momentum tensor (EMT) $T$. This trace is responsible for incorporating supplementary contributions from the quantum realm into conventional gravity. Several researchers have explored cosmology using the $f (Q, T)$ gravity framework \cite{ap, ap0, ap00, ap1, ap2, ap3, ap4, ap5, ap7, ap8, ap9, ap10, ap11, ap12, ap13, ap14, ap15}.

Recently, numerous studies have been conducted on wormhole solutions in the context of $f(Q, T )$ gravity. In \cite{1}, the study examines wormhole solutions within the context of $f(Q,T)$ gravity theory, which includes conformal symmetries. The authors investigate the potential for traversable wormholes across various scenarios, taking into account traceless, anisotropic, and barotropic EoS. Also, conformally symmetric wormhole solutions supported by non-commutative geometries in
the context of $f(Q,T)$ gravity are presented in \cite{11, 12a, 12}.
 Nashed and Hanafy have investigated stable soliton dark matter (DM) wormholes in non-minimally coupled $f(Q, T)$ gravity \cite{2}.
Impact of DM galactic halo models on wormhole geometry within $f(Q,T)$ gravity is investigated in \cite{3}. Sadatian and Hosseini have employed numerical analysis to derive wormhole solutions that satisfy the weak energy condition (WEC) \cite{4}. In \cite{5}, a linear function and a quadratic function of $Q$ have been utilized to explore wormhole solutions. Wormholes with EoS in the form of $p=\omega(\rho-4B)$, where $\omega$ is the EoS parameter and $B$ is the bag parameter, have been studied in \cite{6}.
In \cite{7}, the authors utilized a non-constant redshift function to derive a non-exotic static spherically symmetric thin-shell wormhole solution. The authors in \cite{8} have assumed a logarithmic shape function and examined the situation for various redshift functions. Additionally, for a particular form of energy density, they have derived a shape function.
Tayde et al. have investigated wormhole solutions within the context of $f(Q,T)$ gravity, focusing on two intriguing non-commutative geometries: Gaussian and Lorentzian distributions as described in string theory \cite{9}.
Casimir wormhole with GUP correction in extended symmetric teleparallel gravity is investigated in \cite{10}. Gul et al. have investigated various models of $f(Q,T)$ theory to derive the explicit formulations of matter contents, which are beneficial for the analysis of wormhole structures \cite{13}. Pradhan et al. have explored the phantom DE behavior in Weyl Type $f(Q, T)$ gravity models with observational constraints \cite{14}.

In this paper, we investigate the implications of  $f(Q,T)$ gravity for the construction and analysis of wormholes. We will explore how this modified theory can yield viable wormhole solutions that satisfy ECs and examine the potential of $f(Q,T)$  gravity to contribute to our understanding of some unsolved problems in GR. Through this investigation, we aim to shed light on the interplay between modified gravity theories and the intriguing concept of traversable wormholes, paving the way for new insights into the fabric of the universe.

To the best of our understanding, all suggested wormhole solutions with constant redshift function within the framework of $f(Q,T)$ gravity violate the classical ECs. As a result, the primary goal of this paper is to identify wormhole solutions that conform to these ECs. Following this, we employ a linear function for $f(Q,T)$ to explore solutions that satisfy the ECs. To achieve this aim, we derive the field equations within the context of $f(Q,T)$ gravity, thereby establishing a distinct relationship between the field equations in the GR framework and those in the realm of $f(Q,T)$.

The manuscript is structured into several sections. Section \ref{sec2} provides a concise introduction to the concept of wormholes. Subsequently, we offer a brief overview of the $f (Q,T)$ theory in conjunction with the classical ECs. Section \ref{sec3} discusses the solutions that adhere to the ECs. In conclusion, we present our findings in Sec.\ref{sec4}. Throughout this document, we have utilized the framework of gravitational units, specifically establishing that $c = 8 \pi G = 1$.

\section{Basic formulation} \label{sec2}
The general line element of a wormhole, characterized by static and spherically symmetric properties, is expressed as follows
\begin{equation}\label{1}
ds^2=-U(r)dt^2+\frac{dr^2}{1-\frac{b(r)}{r}}+r^2(d\theta^2+\sin^2\theta,
d\phi^2)
\end{equation}
where $U(r)=\exp2\phi(r)$. Here, $\phi(r)$ is referred to redshift function associated with the gravitational redshift. Also, $b(r)$ is known as the shape function with direct effect on the curvature and the geometry of the wormhole.
The throat of a wormhole links two separate universes or distinct areas of the same universe. The throat condition is
\begin{equation}\label{2}
b(r_0)=r_0
\end{equation}
where $r_{0}$ is the throat of wormhole. The existence of a traversable wormhole depends on the following conditions being met
\begin{equation}\label{3}
b'(r_0)<1
\end{equation}
and
\begin{equation}\label{4}
b(r)<r,\ \ {\rm for} \ \ r>r_0.
\end{equation}
Equation (\ref{3}) is referred to as the flaring-out condition, which indicates a violation of the NEC within the context of GR. Furthermore, the wormhole solutions must satisfy the asymptotically flat condition to be in agreement with the cosmos on a large scale,
\begin{equation}\label{5}
\lim_{r\rightarrow \infty}\frac{b(r)}{r}=0,\qquad   \lim_{r\rightarrow \infty}U(r)=1.
\end{equation}
To ensure that there are no tidal forces, the redshift function must be constant. For the sake of simplicity, we consider the vanishing redshift function with a lot of utilization in the literature. In this work, we focus on the anisotropic fluid described by the tensor $T_{\nu}^{\mu}=diag[-\rho,p,p_{t},p_{t}]$. The energy density is represented by $\rho$, while $p$ and $p_{t}$ show the radial and tangential pressures, respectively.
In continuation, we will exhibit a brief report of the $f(Q,T)$ gravity.
In this context, the action is
\begin{equation}\label{6a}
S=\int\frac{1}{2}\,f(Q,T)\sqrt{-g}\,d^4x+\int L_m\,\sqrt{-g}\,d^4x\, ,
\end{equation}
where $f(Q,T)$ represents a general function of $Q$ and $T$, $\sqrt{-g}$ is the determinant of the metric, and $L_{m}$ specifies the matter lagrangian density\cite{6}. The non-metricity tensor is given by\cite{6}
\begin{equation}\label{6ab}
Q_{\lambda\mu\nu}=\bigtriangledown_{\lambda} g_{\mu\nu},
\end{equation}
In addition, the non-metricity conjugate or superpotential is defined as
\begin{eqnarray}\label{6ab1}
P^\alpha\;_{\mu\nu}=\frac{1}{4}\Bigg[&-&Q^\alpha\;_{\mu\nu}+2Q_{(\mu}\;^\alpha\;_{\nu)}\nonumber \\
&+&Q^\alpha g_{\mu\nu}-\tilde{Q}^\alpha g_{\mu\nu}-\delta^\alpha_{(\mu}Q_{\nu)}\Bigg],
\end{eqnarray}
where
\begin{equation}\label{a1}
Q_{\alpha}=Q_{\alpha}\;^{\mu}\;_{\mu},\; \tilde{Q}_\alpha=Q^\mu\;_{\alpha\mu},
\end{equation}
are two traces of the non-metricity tensor. The non-metricity scalar symbolized as\cite{6}
\begin{eqnarray}\label{a2}
Q &=& -Q_{\alpha\mu\nu}\,P^{\alpha\mu\nu} \nonumber \\
&=& -g^{\mu\nu}\left(L^\beta_{\,\,\,\alpha\mu}\,L^\alpha_{\,\,\,\nu\beta}-L^\beta_{\,\,\,\alpha\beta}\,L^\alpha_{\,\,\,\mu\nu}\right),
\end{eqnarray}
with
\begin{equation}\label{a3}
L^\beta_{\,\,\,\mu\nu}=\frac{1}{2}Q^\beta_{\,\,\,\mu\nu}-Q_{(\mu\,\,\,\,\,\,\nu)}^{\,\,\,\,\,\,\beta}.
\end{equation}
Now, by varying the action with respect to the metric tensor, the field equations are yielded
 \begin{multline}\label{6b}
\frac{-2}{\sqrt{-g}}\bigtriangledown_\alpha\left(\sqrt{-g}\,f_Q\,P^\alpha\;_{\mu\nu}\right)-\frac{1}{2}g_{\mu\nu}f \\
+f_T \left(T_{\mu\nu} +\Theta_{\mu\nu}\right)\\
-f_Q\left(P_{\mu\alpha\beta}\,Q_\nu\;^{\alpha\beta}-2\,Q^
{\alpha\beta}\,\,_{\mu}\,P_{\alpha\beta\nu}\right)= T_{\mu\nu},
\end{multline}
where $f_{Q}=\frac{\partial f}{\partial Q}$ and $f_{T}=\frac{\partial f}{\partial T}$.
The EMT can be written as
 \begin{equation}\label{a4}
T_{\mu\nu}=-\frac{2}{\sqrt{-g}}\frac{\delta\left(\sqrt{-g}\,L_m\right)}{\delta g^{\mu\nu}},
\end{equation}
and
\begin{equation}\label{a5}
\Theta_{\mu\nu}=g^{\alpha\beta}\frac{\delta T_{\alpha\beta}}{\delta g^{\mu\nu}}.
\end{equation}
We have posited that $L_m$ is solely dependent on the metric component and remains uninfluenced by its derivatives.
By incorporating the metric ( \ref{1}) and the anisotropic fluid into the field equations ( \ref{6b}), we derive the following field equations.
\begin{eqnarray}\label{3e}
  \rho =\frac{1}{2 r^3} \Bigg[f_Q \left(\frac{(2 r-b) \left(r b'-b\right)}{(r-b)}+2b\right) \nonumber \\
  +2 b r f_{\text{QQ}} Q'+f r^3+2r^3 f_T (L_m-\rho )\Bigg],
\end{eqnarray}

\begin{eqnarray}\label{3f}
 p=-\frac{1}{2 r^3} \Bigg[f_Q b\left(\frac{r b'-b}{r-b}+2\right) +2 b r f_{\text{QQ}} Q'\nonumber \\
 +f r^3+2r^3 f_T \left(L_m+p_r\right)\Bigg],
\end{eqnarray}
\begin{eqnarray}\label{3g}
  p_t=-\frac{1}{2 r^2} \Bigg[f_Q \left(\frac{\left(r b'-b\right) }{r-b }\right)+f r^2\nonumber \\
  -2r^2 f_T \left(-L_m-p_t\right)\Bigg].
\end{eqnarray}
The ultimate expression of the field equations is contingent upon the functions $f(Q, T)$ and $L_m$. In gravities that are based on the Lagrangian framework, three prevalent selections for $L_m$ are $L_m=-\rho$, $L_m=-T$, and $L_m=-P$ where, $T=-\rho+p+2p_t$ and $P=\frac{p+2p_t}{3}$
We will consider $L_m=-P$ along with a linear model for $f(Q, T)$ in the subsequent section to derive wormhole solutions.

\par The energy conditions (ECs) are mathematically imposed on the solutions of the Einstein field equations to guarantee the existence of the physical solutions. Wormhole solutions in the framework of GR have been shown to violate ECs. These conditions contain the WEC,  NEC, dominant energy condition(DEC), and strong energy condition(SEC),
\begin{eqnarray}\label{21}
\textbf{WEC}&:& \rho+p\geq 0,\quad \rho+p_t\geq 0 \\
\label{21a}
\textbf{NEC}&:& \rho\geq 0, \rho+p\geq 0,\quad \rho+p_t\geq 0, \\
\textbf{DEC}&:& \rho\geq 0, \rho-|p|\geq 0,\quad \rho-|p_t|\geq 0, \\
\textbf{SEC}&:& \rho+p\geq 0,\, \rho+p_t\geq 0,\rho+p+2p_t \geq 0. \label{21b}
\end{eqnarray}
The various energy conditions are not arbitrary prescriptions; rather, they emerge naturally from the Raychaudhuri equation, which dictates the evolution of the expansion scalar for a congruence of timelike or null geodesics. In its most basic form, the Raychaudhuri equation connects the rate of change of the expansion scalar to the Ricci curvature term for a null vector, along with contributions from shear and rotation. By enforcing the physically reasonable condition that gravity must be attractive—ensuring that initially converging geodesics continue to focus—one requires that the Ricci term be non-negative. This condition, through the Einstein field equations, translates into inequalities concerning the stress-energy tensor, resulting in the null energy condition. At this juncture, as cited in \cite{fq}, we will examine the ECs in the following sections of this paper by defining the functions,
\begin{eqnarray}\label{22}
 H(r)&=& \rho+p ,\, H_1(r)= \rho+p_t,\, H_2(r)= \rho-|p|, \nonumber \\
 H_3(r)&=&\rho-|p_t|,\, H_4(r)= \rho+p+2p_t .
\end{eqnarray}
Researchers have employed various strategies to discover asymptotically flat wormhole solutions in the framework of GR or other modified theories. The primary method involves analyzing an EoS to derive the shape function through the application of field equations. In contrast, some researchers utilize shape functions characterized by free parameters, which they subsequently modify to pinpoint solutions that meet both physical and mathematical criteria. In the subsequent section, we will apply these methods to find asymptotically flat wormhole solutions. Here, for the sake of simplicity, we put $r_{0}=1$ in the continuation of this work.

\section{Non-exotic wormhole solutions }\label{sec3}
In this section, we adopt a linear form for $f(Q,T)$ as follows
\begin{equation}\label{99a}
f(Q, T)=\alpha  Q+\beta T,
\end{equation}
where $\alpha$ and $\beta$ are free parameters. This model was introduced in \cite{fqt} and naturally describes
an exponentially expanding universe. Using this function in Eqs.(\ref{3e}-\ref{3g}) gives the following field equation
\begin{equation}\label{18}
\rho =A_1\frac{ b'}{ r^2},
\end{equation}
\begin{equation}\label{19}
p= \frac{A_2 b+A_3 rb'}{r^3},
\end{equation}
\begin{equation}\label{20}
p_t= \frac{A_4 b + A_5r b'}{r^3},
\end{equation}
where

\begin{eqnarray}\label{19aa}
A_1&=&\frac{\alpha\left(\frac{3}{2}-\beta\right)}{3(1+\beta)(\frac{1}{2}-\beta)},\nonumber \\
A_2&=&-\frac{\alpha}{1+\beta}=-2A_4,\nonumber \\
A_3&=&-\frac{2\alpha \beta}{3(1+\beta)(\frac{1}{2}-\beta)},\nonumber \\
A_5&=&\frac{\alpha\left(\frac{3}{2}+\beta\right)}{6(1+\beta)(\frac{1}{2}-\beta)}.
\end{eqnarray}
It is easy to show that
\begin{eqnarray}\label{11cc}
H(r)&=&\rho(r)+p(r)\nonumber \\
&=&\gamma(\alpha,\beta) \frac{rb'-b}{r^3},
\end{eqnarray}
and
\begin{eqnarray}\label{11d}
H_1(r)&=&\rho(r)+p_t(r)\nonumber \\
&=&\gamma(\alpha,\beta) \frac{rb'+b}{2r^3},
\end{eqnarray}
where
\begin{equation}\label{2a}
\gamma(\alpha,\beta)=\frac{\alpha}{1+\beta}=-A_2.
\end{equation}
Equations (\ref{11cc}) and (\ref{11d}) suggest that
\begin{eqnarray}\label{1cc}
H^f&=&\gamma(\alpha,\beta)H^G\nonumber \\
H_1^f&=&\gamma(\alpha,\beta)H_1^G,
\end{eqnarray}
where the indic $f$ pertains to the $f(Q,T)$ gravity and the indic $G$ is employed in the GR formalism. Therefore, it is clear that solutions which concurrently breach both radial and lateral NEC within the framework of GR can still respect to the NEC in $f(Q,T)$ theory when
 \begin{equation}\label{a222}
\gamma(\alpha,\beta)<0,
\end{equation}
 is satisfied.
 It can be inferred that the result of this model aligns with $f(R, T)=\alpha R+ \beta T$, within the framework of $f(R, T)$ theory, which was previously analyzed in Ref. \cite{SR1} and $f(R,L_m)=\alpha R+ \beta L_m$, in the context of $f(R, L_m)$ theory \cite{SR2} as well as $f(T,\mathcal{T})$ gravity \cite{foad3}.

 We will utilize this significant aspect to investigate non-exotic wormhole solutions in this paper. It is easy to  show that in the range
\begin{equation}\label{cc1}
\alpha>0,\quad\beta<-1
\end{equation}
or
\begin{equation}\label{cc2}
\alpha<0,\quad\beta>-1,
\end{equation}
$\gamma$ is negative.
We currently possess multiple strategies for addressing the field equations. We have three equations, Eqs. (\ref{18}),(\ref{19}) and (\ref{20}), which involve four unknown functions: $b(r)$, $ \rho(r)$, $p(r)$, and $p_t(r)$. To complete the system, one might consider an additional EoS given by $p(\rho)=F(\rho)$, or one could explore specific selections for the distribution of the energy density of the wormhole, as referenced in \cite{2, 3}. Another approach, as utilized in \cite{4, 6}, involves manually specifying the redshift function and the shape function, denoted as $ \phi(r)$ and $b(r)$, respectively. In the following section, we will adopt two EoS and seek solutions that satisfy the ECs.

\subsection{ Solutions with linear EoS}\label{subsec1}
As the first model, we consider a linear EoS as follows
\begin{equation}\label{9a}
 p(r)=\omega \rho(r).
\end{equation}
where $\omega$ is the constant EoS parameter. This relationship holds significant importance for the modeling of different types of matter, such as perfect fluids, DE, and specific phases of matter present in the early universe. Furthermore, the linear EoS acts as a fundamental model for more intricate situations, facilitating a more profound comprehension of the interaction between gravitational influences and thermodynamic characteristics. Using the EoS (\ref{9a}) in field equations (\ref{18}) and (\ref{19}) gives
\begin{equation}\label{18a}
b(r)=r^{n(\omega, \beta)},
\end{equation}
where
\begin{equation}\label{18b}
n(\omega, \beta)=\frac{3(1-2\beta)}{2\omega \beta-4\beta-3\omega}.
\end{equation}
The power-law shape function is the most renowned shape function within the realm of wormhole theory. It  meets all the requirements for wormhole theory as long as $n<1$ is true. Let us explore the ECs for this shape function. Using Eq.(\ref{18}) and the shape function (\ref{18a}) implies that the condition
\begin{equation}\label{18bb}
  n(\omega, \beta)A_1>0,
 \end{equation}
must hold to have a positive energy density. It is easy to show that
\begin{equation}\label{a2a}
A_1=\gamma \frac{\frac{3}{2}-\beta}{3\left(\frac{1}{2}-\beta\right)}.
\end{equation}
\begin{figure}
\subfloat[ $n>1$ ]{\includegraphics[width = 1.9in]{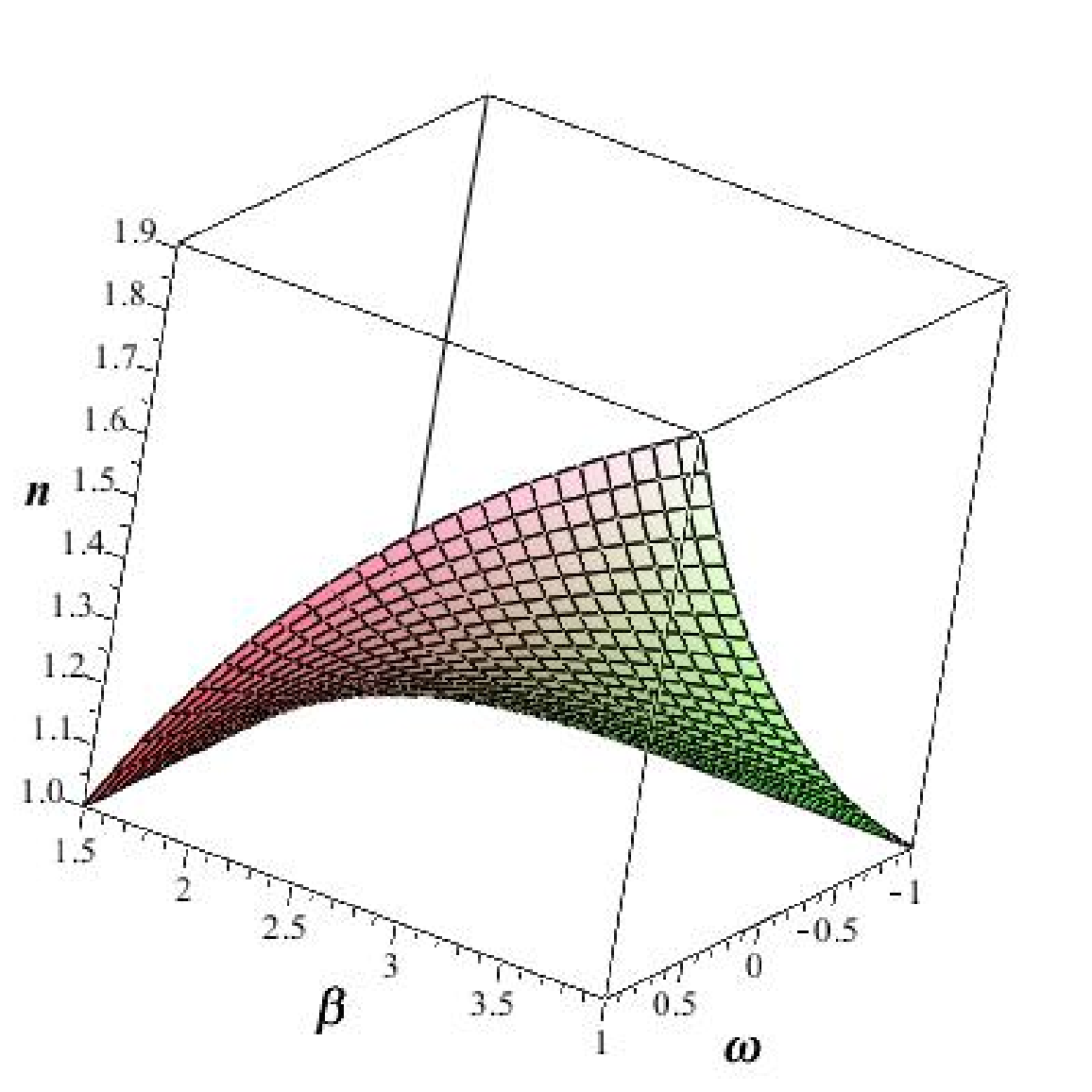}}\\
\subfloat[ $0<n<1$]{\includegraphics[width = 1.9in]{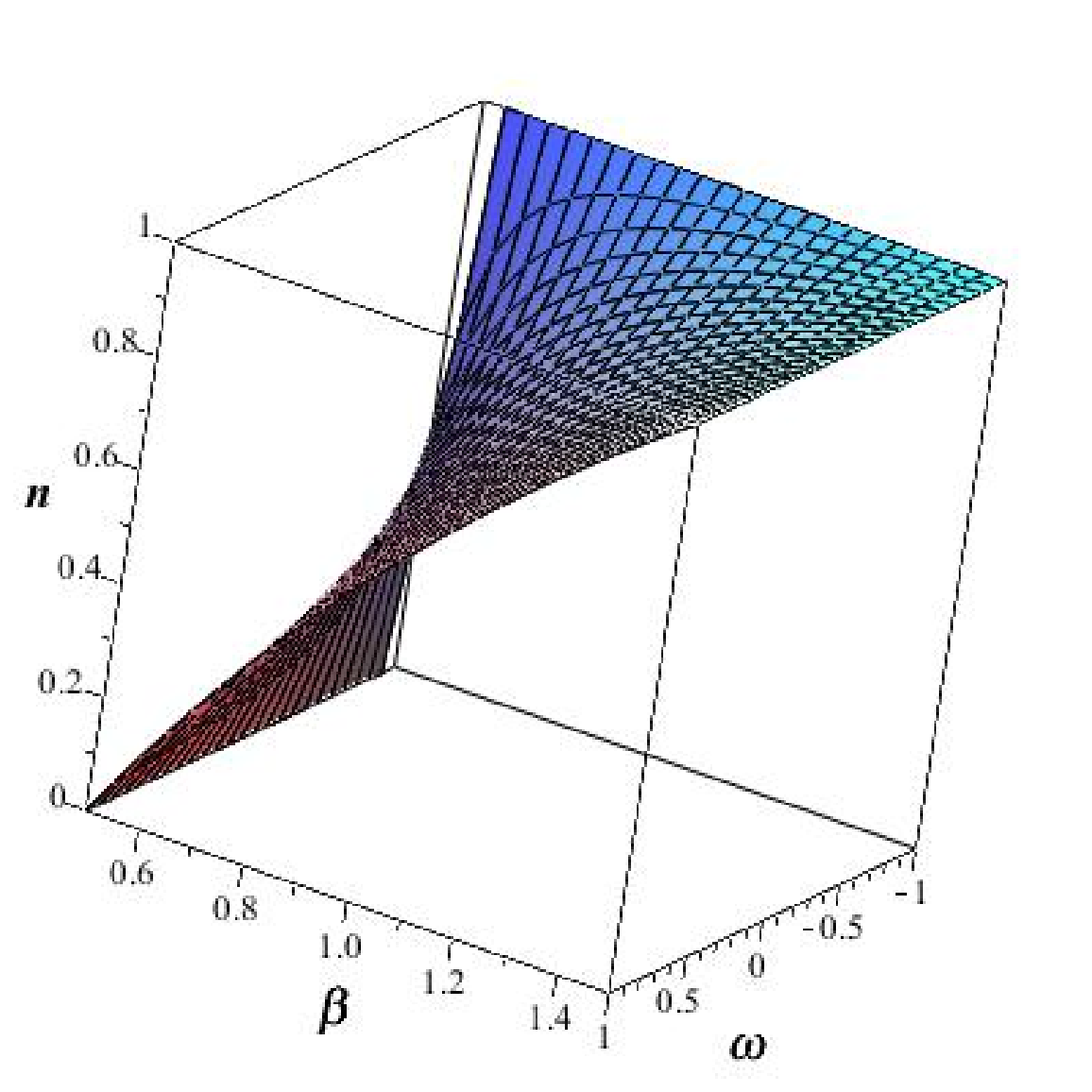}}\\
\subfloat[ $n<0$ ]{\includegraphics[width = 1.9in]{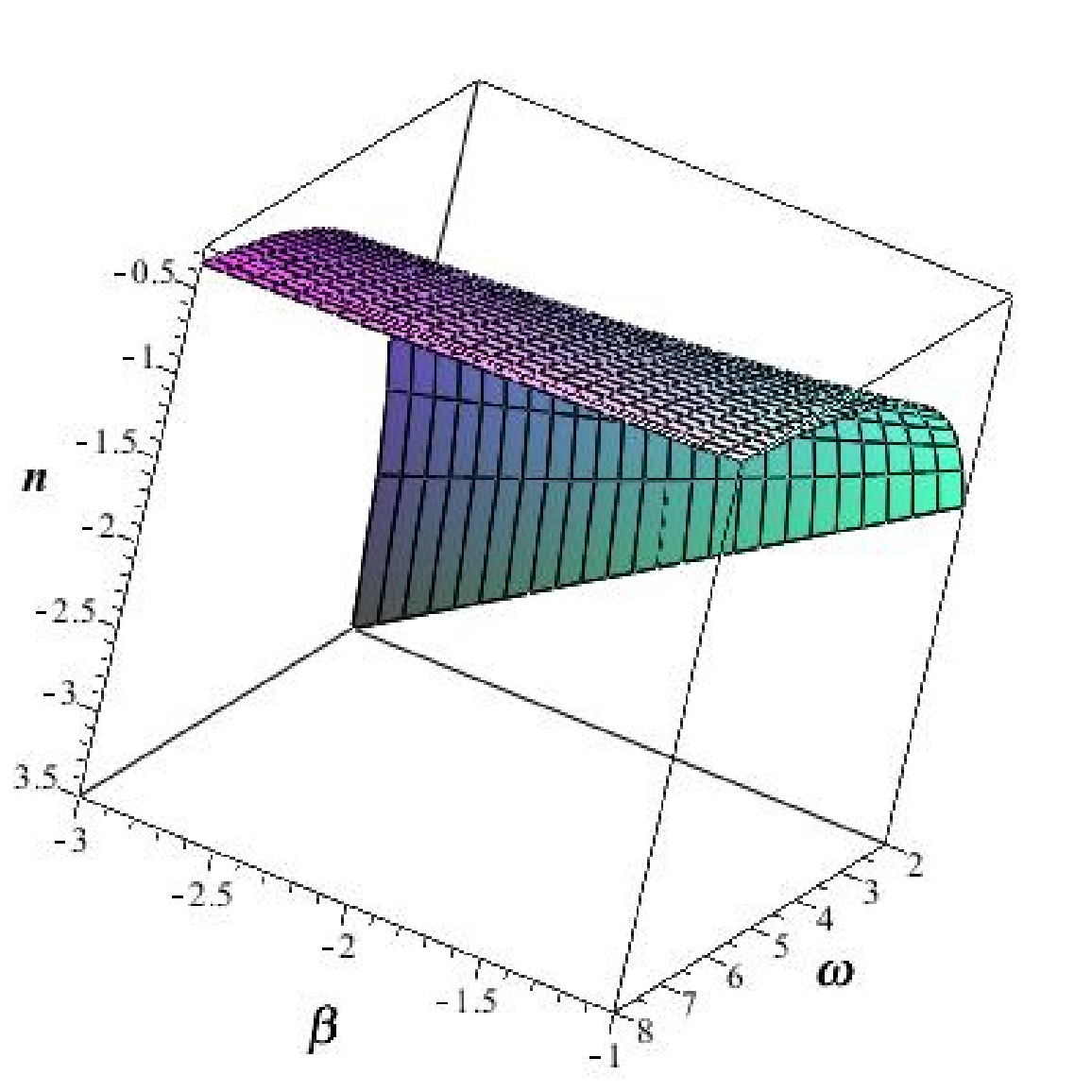}}
\subfloat[ $n<0$]{\includegraphics[width = 1.9in]{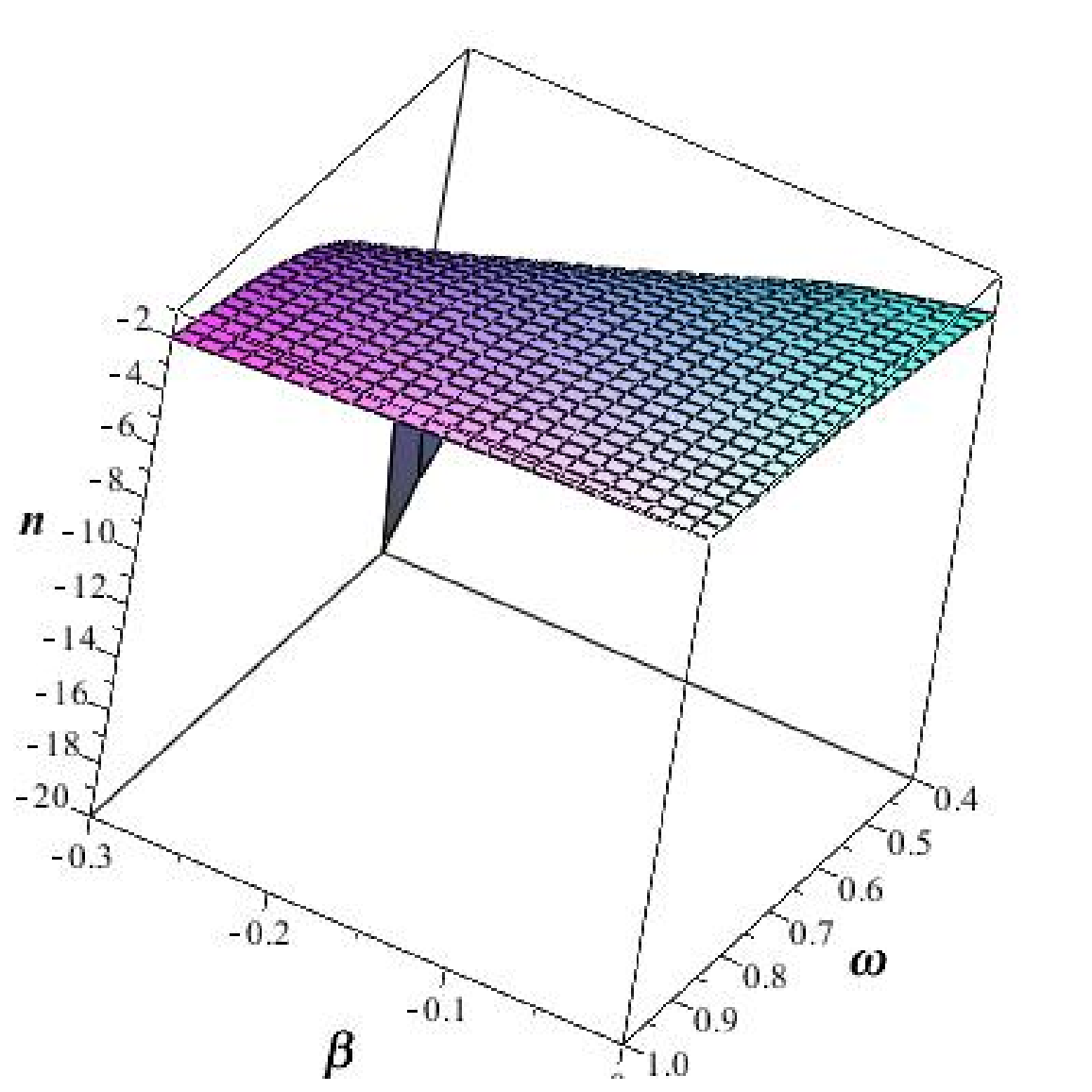}}
\caption{The graph depicts the correlation between $n(\omega, \beta)$ and the variables $\omega$ and $\beta$. It is clear that $n(\omega, \beta)$ is positive within a certain range of $\alpha$ and $m$ (a, b) , , while $n$ is negative in some different ranges (c, d).}
\label{fig1}
\end{figure}
\begin{table*}
\begin{tabular}{|l|l|l|l|l|l|l|l|}
\hline
$b(r)$  & $\rho>0$ & $H>0$  & $H_1>0$ & $H_2>0$ & $H_3>0$ & $H_4>0$  & Satisfied ECs  \\ \hline
 $r^{6/7}$  & \checkmark & \checkmark & $\times$ & \checkmark & $\times$ & $\times$ &   \\ \hline
 $r^{2/3}$  & \checkmark & \checkmark & $\times$ & \checkmark & $\times$ & $\times$ &   \\ \hline
$r^{-15/34}$ &  \checkmark & \checkmark & $\times$ & $\times$ & $\times$   & \checkmark  & \\ \hline
$r^{-33/19}$ &  \checkmark & \checkmark & \checkmark & $\times$ & \checkmark   & \checkmark  & $WEC$, $NEC$, $SEC$ \\ \hline
 $r^{-6}$ & \checkmark & \checkmark & \checkmark & \checkmark & \checkmark & \checkmark & everyone \\ \hline
\end{tabular}
\caption{The results of ECs for some individual $b(r)$.  }\label{Tab1}
\end{table*}

Given that $\gamma<0$ is a crucial requirement for the existence of non-exotic wormhole solutions, it is possible to examine the sign of
\begin{equation}\label{24a}
\frac{\frac{3}{2}-\beta}{3\left(\frac{1}{2}-\beta\right)},
\end{equation}
for the purpose of verifying condition (\ref{18bb}). It can be readily demonstrated that the term (\ref{24a}) is negative within the interval
\begin{equation}\label{241}
\frac{1}{2}<\beta<\frac{3}{2}
\end{equation}
and positive outside of it. To ensure the validation of the radial NEC for a positive energy density, it is required that $-1<\omega$. We have illustrated the function $n(\omega, \beta)$ in relation to $\omega$ and $\beta$ in Fig.(\ref{fig1}) across various ranges. It is clear that $n(\omega, \beta)<1$ is permissible within certain ranges of $\omega$ and $\beta$. We have selected five options based on the assessable ranges of $\omega$ and $\beta$ as shown in Fig.(\ref{fig1}). In the first case, we set $\alpha=-\beta=2\omega=-1$, resulting in $b(r)=r^{(6/7)}$. The second choice is $\alpha=-\beta=-2\omega=-1$, which leads to $b(r)=r^{(2/3)}$. The third case involves $\alpha=4\beta=-2\omega=-1$, resulting in $b(r)=r^{(-6)}$. The fourth one is  $\alpha=-\frac{\beta}{2}=\frac{\omega}{6}=1$, resulting in $b(r)=r^{(-15/34)}$.  The final choice is $\alpha=-\frac{\beta}{5}=\frac{\omega}{3}=1$, which gives $b(r)=r^{(-33/19)}$. We have plotted $\rho(r)$, $H(r)$, $H_1(r)$, $H_2(r)$, $H_3(r)$, and $H_4(r)$ for these five selections. The results are succinctly displayed in table (\ref{Tab1}). This table illustrates that while the power-law shape function is the most renowned shape function in wormhole theory, many of the selections for this shape function contravene the ECs. It is only through the precise adjustment of the free parameters $\alpha$, $\beta$, and $\omega$ that we can identify the solutions that comply with the ECs.

\subsection{ Asymptotically linear EoS }\label{subsec2}
\begin{figure}
\centering
  \includegraphics[width=3 in]{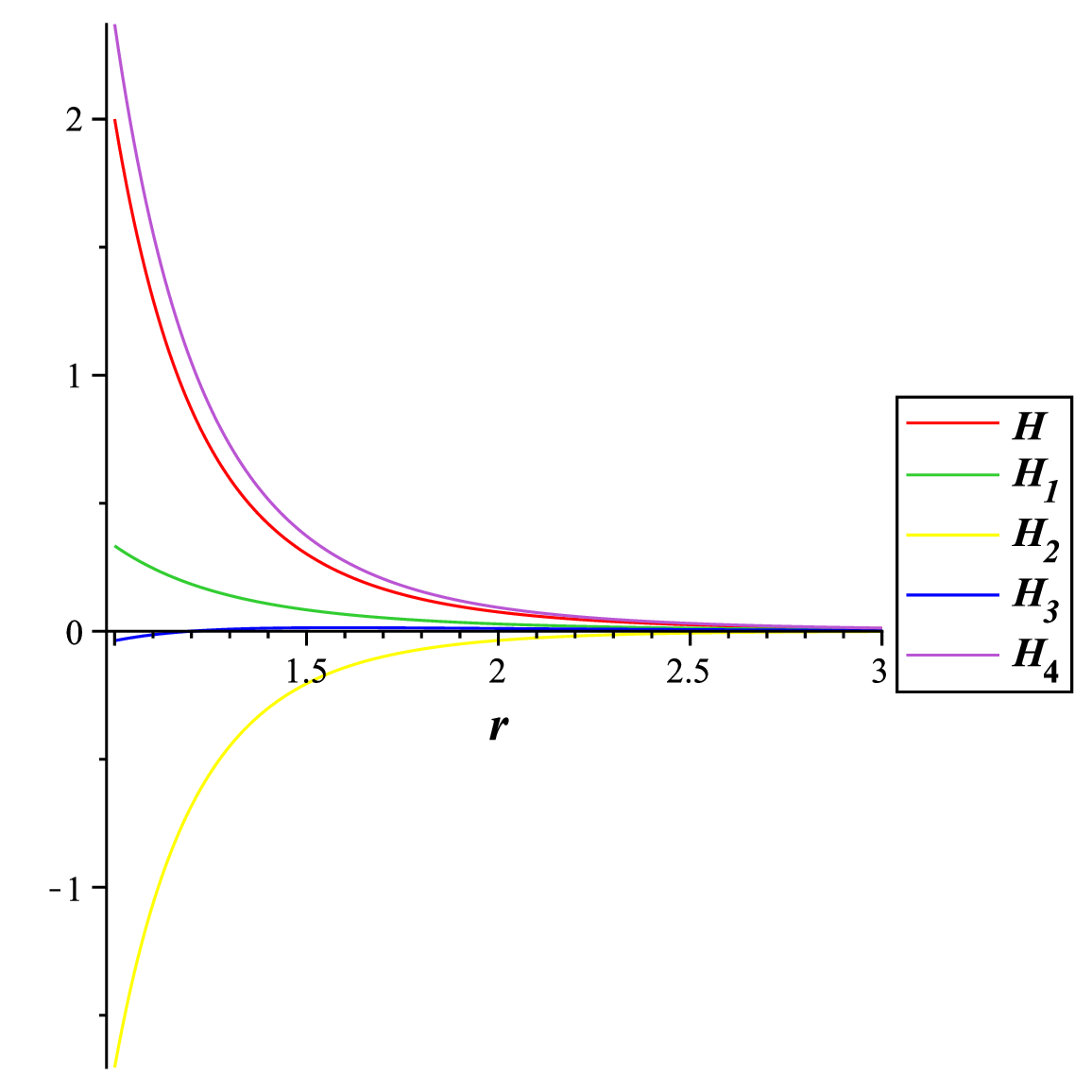}
\caption{The graph depicts the functions $ H(r)$(red), $H_1(r)$(green), $H_2(r)$(yellow), $H_3(r)$(blue), and $H_4(r)$(violet) plotted against the radial coordinate for the shape function $b(r)=\left(\frac{2}{r+1}\right)^4$ within the framework of $f(Q, T)$. It is evident that all ECs, except for the DEC, are satisfied. See the text for details.}
 \label{fig2}
\end{figure}

\begin{figure}
\centering
  \includegraphics[width=3 in]{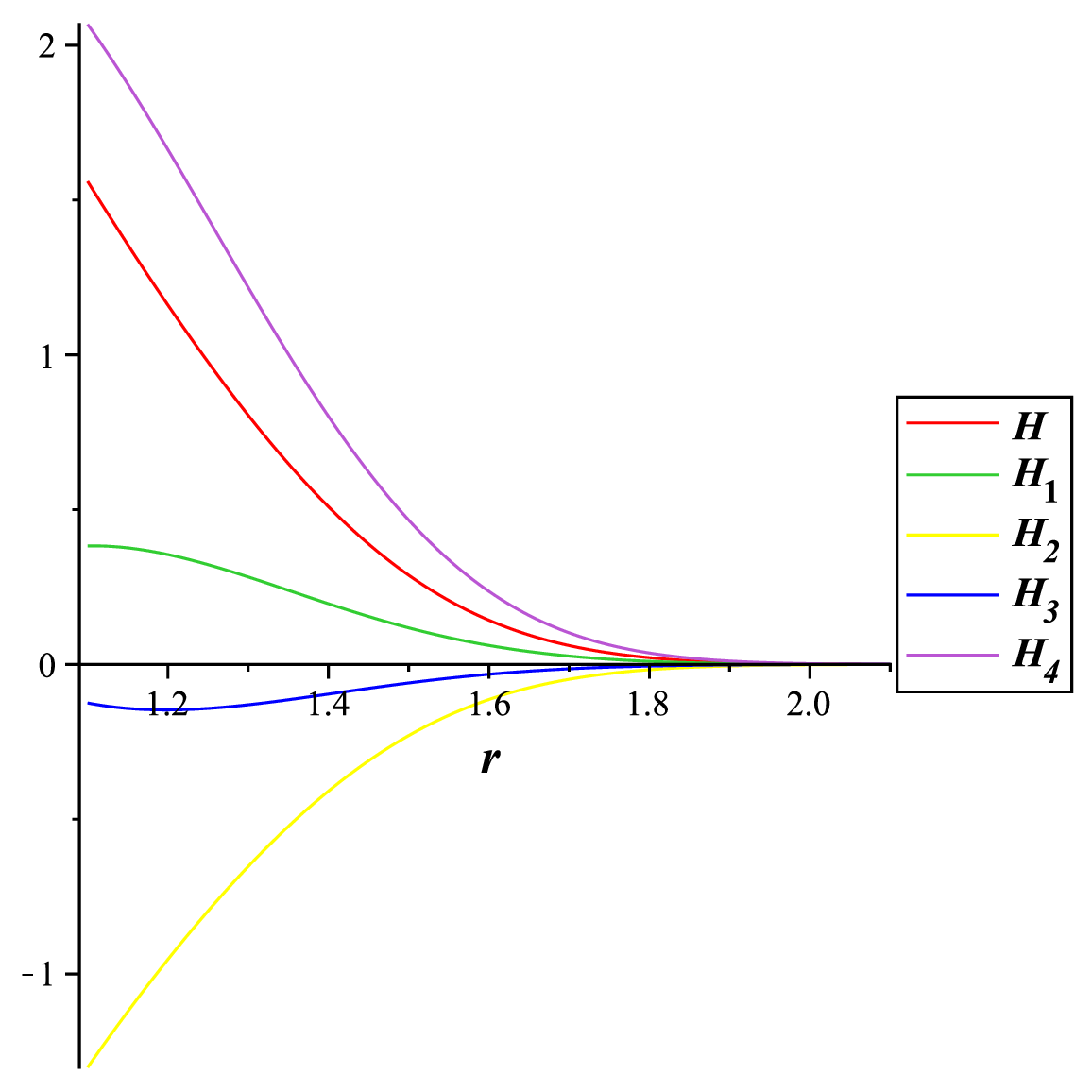}
\caption{The graph depicts the functions $ H(r)$(red), $H_1(r)$(green), $H_2(r)$(yellow), $H_3(r)$(blue), and $H_4(r)$(violet) plotted against the radial coordinate for the shape function $b(r)=\exp(\frac{1-r^4}{2})$ within the framework of $f(Q, T)$. It is evident that all ECs, except for the DEC, are satisfied. See the text for details.}
 \label{fig3}
\end{figure}

The linear EoS is the most straightforward and commonly used EoS found in literature to explain the cosmos and wormhole theory. However, we have recently demonstrated that a wider range of solutions can be explored within the framework of wormhole theory using variable EoS \cite{variable}. This approach is based on the premise that asymptotically linear EoS can be regarded as a more general form of EoS rather than a strictly linear one, as it appears more physically plausible to consider the cosmos as behaving like a perfect fluid on a global scale rather than on a local scale. We have used this method to find solutions in the context of GR \cite{variable}, $f(R,T)$ gravity \cite{SR1}, $f(T,\mathcal{T})$ gravity \cite{foad3}, and $f(R, L_m)$ gravity \cite{SR2}. The same algorithm can be used to find asymptotically flat wormhole solutions in $f(Q,T)$ gravity. Now, we consider a variable EoS as follows
\begin{equation}\label{28bb}
p(r)=\omega(r)\rho(r),
\end{equation}
where
\begin{equation}\label{28bc}
\omega(r)=\omega_\infty+g(r).
\end{equation}
Here, $\omega_\infty$ is the EoS parameter at large radial distance from the throat and $g(r)$ must obey
\begin{equation}\label{28bd}
\lim_{r\longrightarrow \infty}g(r)=0.
\end{equation}
By using (\ref{28bb}) in (\ref{18}) and (\ref{19}), one can find
\begin{equation}\label{38}
b(r)=C \exp\left(\int \frac{A_2 \,\,dr}{r\left(-A_3+(\omega_\infty+g(r))A_1\right)}\right),
\end{equation}
where $C$ is the constant of integration. We will use condition (\ref{2}) to find this constant. In \cite{SR1}, various $g(r)$ functions have been employed to identify solutions that contravene simultaneity regarding radial and lateral NEC within the framework of GR, which subsequently allows for NEC in the context of $f(R,T)$. We can use the same procedure to find the non-exotic wormhole solutions. On the other side, Eq.(\ref{1cc}) explains that all of the solutions that admits the NEC in the context of $f(R,T)$ in \cite{SR1}, may still respect the NEC in the context of $f(Q,T)$. The only point that should be considered in the mind is the positivity of  $\rho(r)$. It was shown in \cite{SR1} that energy density is as follows
 \begin{equation}\label{39a}
\rho(r)=\frac{1}{1+2\lambda}\frac{b'}{r^2}.
\end{equation}
Since in the $f(R,T)$ scenario $1+2\lambda$ is negative, therefore to have a positive energy density, the condition
\begin{equation}\label{39}
\frac{b'}{r^2}<0
\end{equation}
must hold. Using (\ref{18}) with (\ref{a222}) and (\ref{39}) explains that
\begin{equation}\label{39b}
\frac{A_1}{\gamma}=\frac{\left(\frac{3}{2}-\beta\right)}{3\left(\frac{1}{2}-\beta\right)}>0,
\end{equation}
is crucial for achieving a positive energy density in the solutions outlined in \cite{SR1}. In the previous subsection, it was mentioned that the condition (\ref{39b}) holds in the region outside of the range, $\frac{1}{2}<\beta<\frac{3}{2}$. As an example, we use two shape functions, previously introduced in \cite{SR1}, and check their physical properties in the context of $f(Q,T)$ gravity. The first example is
\begin{equation}\label{a22}
b(r)=(B+D)^{\frac{1}{B}}(Br+D)^{-\frac{1}{B}}.
\end{equation}
 This shape function violates radial and lateral NEC in the context of GR  making it a potential candidate for fulfilling the NEC in the context of $f(Q,T)$ gravity. It was shown that
 \begin{equation}\label{a23}
0<B\leq 1-D
\end{equation}
 must hold to satisfy the ECs in the context of $f(R.T)$ \cite{SR1}. In the $f(Q,T)$ scenario, we set $\alpha=-\beta=-8D=-8B=-2$ which gives
 \begin{equation}\label{f30}
b(r)=\left(\frac{2}{r+1}\right)^4
\end{equation}
 so
\begin{equation}\label{f31}
\rho(r)=\frac{128}{27r^2(r+1)^5}.
\end{equation}
The energy density is positive so we have plotted $H$, $H_1$, $H_2$, $H_3$, and $H_4$ against radial coordinate in Fig. (\ref{fig2}). This figure implies that all of the ECs except the DEC are respected in the context of $f(Q,T)$ for the shape function (\ref{f30}). Now, we can find the EoS parameter as follows
\begin{equation}\label{39aa}
\omega(r)=\omega_\infty+g(r)=\frac{p(r)}{p_t(r)}=\frac{41r+9}{4r}.
\end{equation}
It can be observed that $\omega_\infty=41/4$ and $g(r)=\frac{9}{4r}$, which correspond to $\omega_\infty=B$ and $g(r)=\frac{D}{r}$ in the context of $f(R,T)$ gravity.  The second example is
\begin{equation}\label{s1}
b(r)= \exp(\frac{1}{nD}-\frac{r^n}{nD}).
\end{equation}
which results from  $\omega_\infty=0$ and $g(r)=\frac{D}{r^n}$ in the context of $f(R,T)$.  This shape function fulfills all the ECs in the realm of $f(R,T)$ gravity \cite{SR1}. We will now analyze it within the framework of $f(Q,T)$ gravity. For example, we set $\alpha=-\beta=-4D=-\frac{n}{2}=-2$ which gives
\begin{equation}\label{8b1}
b(r):=\exp(\frac{1}{2}-\frac{r^4}{2})
\end{equation}
Using (\ref{8b1}) in (\ref{18}) leads to
\begin{equation}\label{29}
\rho=\frac{4r}{27}\exp(\frac{1}{2}-\frac{r^4}{2}).
\end{equation}
%\begin{equation}\label{a29}
%p=\frac{\rho\left( (\alpha-1)r^3 -b \rho^{-\alpha} \right)}{\alpha r^3}.
%\end{equation}
%\begin{equation}\label{25}
% p_t=\frac{\rho^{(1-\alpha)}\left( b-r b' +2(\alpha-1) r^3  \rho^\alpha \right)}{2\alpha r^3}.
%\end{equation}
The energy density is positive, so we have illustrated $H$, $H_1$, $H_2$, $H_3$, and $H_4$ in relation to the radial coordinate in Fig. (\ref{fig3}). This figure indicates that all of the ECs, with the exception of the DEC, are satisfied within the framework of $f(Q,T)$ for the shape function (\ref{8b1}). It is easy to demonstrate that the shape function (\ref{s1}) gives
\begin{equation}\label{209b}
\omega(r)=\frac{4\beta}{2\beta-3}+\frac{3d(2\beta-1)}{(2\beta-3)r^n},
\end{equation}
which is different from $f(R,T)$ gravity.

\begin{figure}
\centering
  \includegraphics[width=3 in]{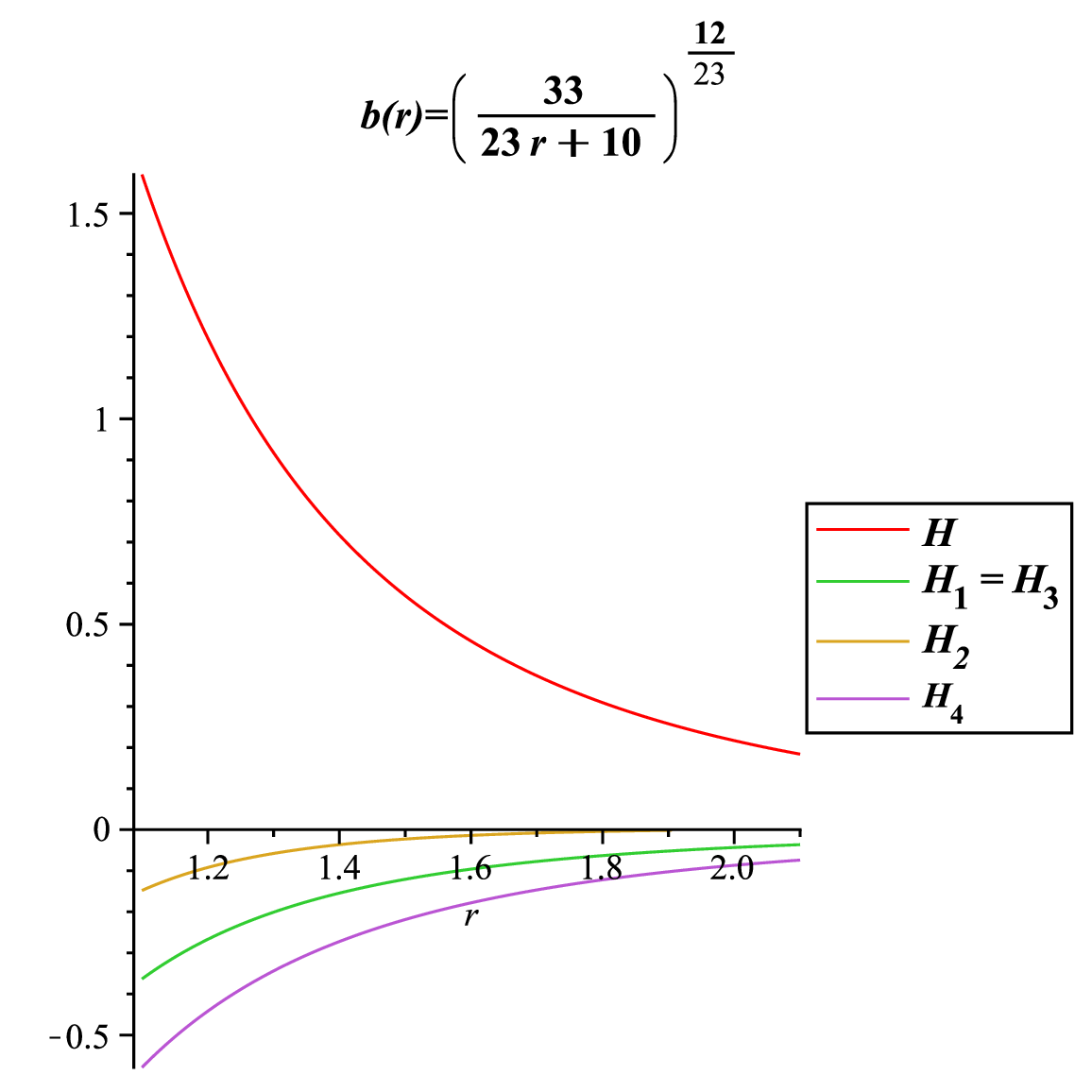}
\caption{The graph depicts the functions $ H(r)$(red), $H_1(r)=H_3(r)$(green), $H_2(r)$(yellow), and $H_4(r)$(violet) plotted against the radial coordinate for the shape function $b(r)=\left(\frac{33}{23r+10}\right)^{\frac{12}{23}}$ within the framework of $f(Q, T)$. It is evident that all ECs are violated. See the text for details.}
 \label{fig4}
\end{figure}

\begin{figure}
\centering
  \includegraphics[width=3 in]{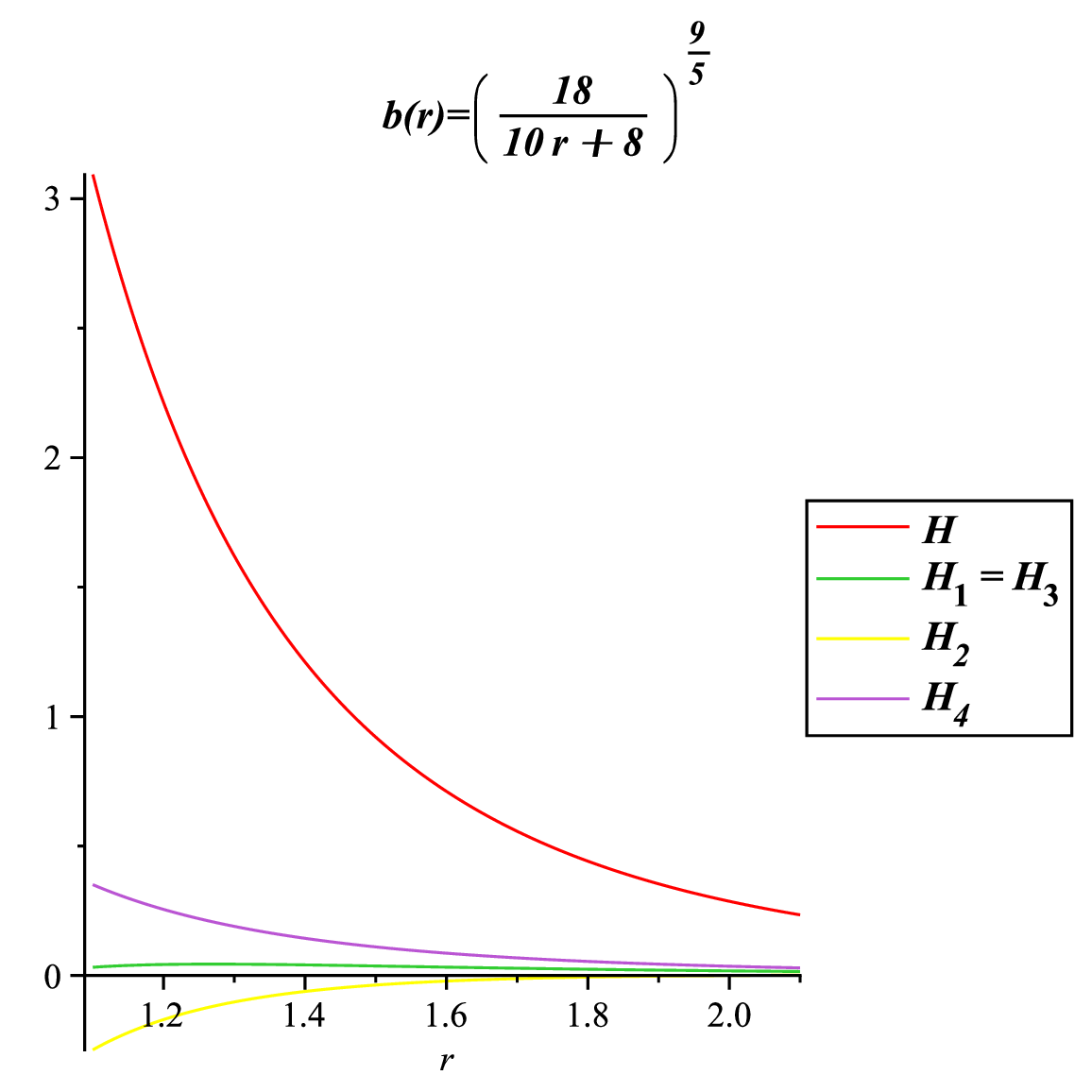}
\caption{The graph depicts the functions $ H(r)$(red), $H_1(r)=H_3(r)$(green), $H_2(r)$(yellow), and $H_4(r)$(violet) plotted against the radial coordinate for the shape function $b(r)=\left(\frac{18}{10r+8}\right)^{\frac{9}{5}}$ within the framework of $f(Q, T)$. It is evident that all ECs, except for the DEC, are satisfied. See the text for details.}
 \label{fig5}
\end{figure}

Let us investigate the solutions with known EoS. As an example, we consider
\begin{equation}\label{s6}
\omega(r)=\omega+\frac{D}{r^n}.
\end{equation}
Using (\ref{s6}) in (\ref{38}) leads to
\begin{equation}\label{s7}
b(r)=(B_1r+D_1)^{n(\omega,\beta)}(B_1+D_1)^{-n(\omega,\beta)}
\end{equation}
which
\begin{eqnarray}\label{s71}
B_1=2\omega\beta-4\beta-3\omega,\nonumber \\
D_1=8\beta-12,
\end{eqnarray}
and $n(\omega,\beta)$ is defined in (\ref{18b}).
Investigating the shape function (\ref{s7}) in its general form presents challenges; therefore, we opt for specific cases of the parameters to finalize the procedure.  Using $\alpha=2\beta=-n/2=4d=-1$ leads to
\begin{equation}\label{s8}
b(r)=\left(\frac{23}{23r+10}\right)^{\frac{12}{23}},
\end{equation}
which gives
\begin{equation}\label{s9}
\rho(r)=\frac{D}{r}.
\end{equation}
We have plotted  $H(r)$, $H_1(r)$, $H_2(r)$, $H_3(r)$ and $H_4(r)$ as a function of $r$ for the shape function (\ref{s8}) in Fig.(\ref{fig4}), which shows all of the ECs are violated. As the second case, we set $\alpha=20\beta=-\frac{8}{3}\omega=-4D=-2$ which leads to
\begin{equation}\label{s99}
b(r)=\left(\frac{18}{10r+8}\right)^{\frac{9}{5}},
\end{equation}
and
\begin{equation}\label{s10}
\rho(r)=\left(\frac{18}{10r+8}\right)^{\frac{14}{15}}\frac{640}{324r^2}.
\end{equation}
\begin{figure}
\centering
  \includegraphics[width=3 in]{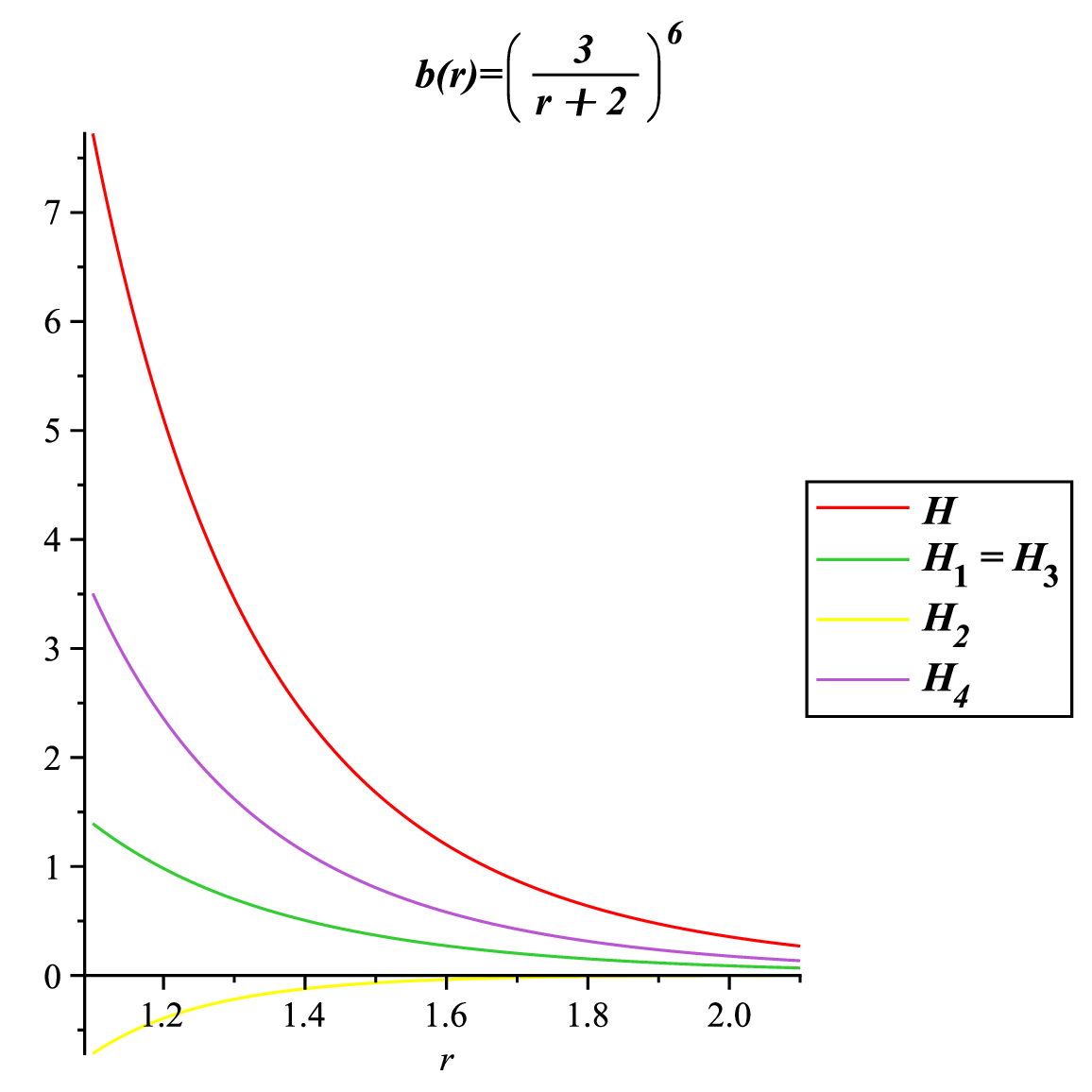}
\caption{The graph depicts the functions $ H(r)$(red), $H_1(r)=H_3(r)$(green), $H_2(r)$(yellow), and $H_4(r)$(violet) plotted against the radial coordinate for the shape function $b(r)=\left(\frac{3}{r+2}\right)^{6}$ within the framework of $f(Q, T)$. It is evident that all ECs, except for the DEC, are satisfied. See the text for details.}
 \label{fig6}
\end{figure}

In Fig.(\ref{fig5}),  $H(r)$, $H_1(r)$, $H_2(r)$, $H_3(r)$ and $H_4(r)$  are depicted as functions of $r$ for shape function (\ref{s99}). This figure illustrates that the DEC is breached in this case; however, other ECs are satisfied. The final choice is $\alpha=-4\beta=-\frac{8}{3}\omega=-4D=-2$ which gives
 \begin{equation}\label{s11}
b(r)=\left(\frac{3}{r+2}\right)^{6},
\end{equation}
and
\begin{equation}\label{s13}
rho(r)=\frac{11664}{(r+2)^7r^2}.
\end{equation}
The results for the shape function (\ref{s11}) are illustrated  in Fig.(\ref{fig6}). This figure illustrates that all of the ECs, with the exception of DEC, are fulfilled for this particular solution. Multiple solutions can be derived by employing different values for the parameters  $\alpha$, $\beta$, $\omega$, and $D$. It can be inferred that a precise adjustment of the parameters may result in solutions that satisfy the ECs. The solutions exhibit a high sensitivity to the free parameters. It is evident that $\beta=-\frac{1}{10}$ in the shape function (\ref{s99}) contrasts with $\beta=-\frac{1}{2}$ in the shape function (\ref{s11}), leading to entirely distinct solutions.

\section{Concluding remarks}\label{sec4}

A wormhole represents a theoretical solution to Einstein's field equations that has yet to be observed experimentally. Stable and traversable wormhole solutions are known to encounter a pathology where ECs are breached. Within the context of GR, the flaring-out condition, which is crucial to wormhole physics, necessarily infringes upon the NEC. Consequently, the requirement for a wormhole in GR entails the presence of exotic matter. As previously noted, modified theories of gravity offer an alternative approach to addressing the challenges associated with wormholes. This study demonstrates that a modified theory of gravity can yield stable solutions without violating ECs.

The modified $f(Q,T)$ gravity, which is based on non-metricity, can be regarded as a non-minimal interaction between matter and geometry. This theory is an extension of symmetric teleparallel gravity. It has proven to be quite effective in elucidating the phenomenon of late-time acceleration and addressing various challenges in cosmology.
We have explored certain theoretical dimensions of the third geometric framework of gravity, known as symmetric teleparallel gravity or $f(Q)$ gravity, by employing a novel class of theories in which the $Q$ is non-minimally coupled to $T$. In this research, we have investigated wormhole models under the linear formulation of $f(Q,T)$ gravity. We utilize a particular representation of the linear model for $f(Q,T)$, defined as $f(Q,T)= \alpha Q + \beta T$, with $\alpha$ and $\beta$ acting as free parameters. It was shown that $\frac{\alpha}{1+\beta}<0$ is the necessary condition to find non-exotic wormhole solutions in the background of $f(Q,T)$ gravity. We have shown that the non-exotic solutions for the linear EoS are only possible in the range (\ref{cc1}) or (\ref{cc2}).

We perform our analysis in two sections: Firstly, wormhole solutions with a linear EoS, and secondly, wormholes with asymptotically linear EoS. Among the different types of EoS, the linear EoS has received considerable attention because of its straightforwardness and relevance in various astrophysical scenarios. Employing a linear EoS results in a power-law shape function, which is the most well-known in the wormhole theory.  Table (\ref{Tab1}) illustrates that out of numerous potential selections for the free parameters, only a few specific instances yield solutions that comply with the ECs.

In the second category of solutions, we have examined that solutions which comply with the ECs in the framework of $f(R,T)$ gravity may also satisfy the ECs in the context of $f(Q,T)$ gravity. The positivity of energy density is a crucial aspect that must be taken into account for this class of solutions. By applying the condition of positive energy density, we can identify non-exotic solutions that were previously presented in the realm of $f(R,T)$ gravity. Although the same shape function in both $f(R,T)$ gravity and $f(Q,T)$ gravity may adhere to the ECs, the physical characteristics of the solutions differ. Another class of solutions is presented by considering a known variable EoS. By using (\ref{38}) a new shape function in the form of (\ref{s7}) is derived. The shape functions, we have presented, meet all the necessary criteria for traversable wormholes, including the throat condition, flaring out condition, and asymptotic flatness condition. It is essential to emphasize that the parameters of the model in question significantly affect the analysis of wormhole shapes. Our research has uncovered the impact of the model parameters $\alpha$ and $\beta$ on the shape functions and the non-exotic property of solutions.

The direct relationship between nonmetricity and matter adds extra degrees of freedom that alter the equations governing the gravitational field and create effective stresses that can sustain wormhole throats without the need for fundamentally exotic sources of matter. As a result, wormholes in $f(Q,T)$ gravity offer significant understanding of how the couplings between matter and geometry affect the topology of spacetime, energy conditions, and the feasibility of traversable solutions within a more extensive gravitational context.


\begin{thebibliography}{99}



%\bibitem{flamm}L. Flamm, Phys. Z. {\bf17}, 448 (1916).

%\bibitem{Rosen} A. Einstein, N. Rosen, Phys. Rev. {\bf48} , 73 (1935).

\bibitem{wheeler}C. W. Misner and J. A. Wheeler, Ann. Phys. {\bf2}, 525 (1957).

\bibitem{WH} M. S. Morris, K. S. Thorne, Am. J. Phys. {\bf 56}, 395 (1988).
%\bibitem{Ellis}H. G. Ellis, J. Math. Phys. (N.Y.) {\bf14} 104 (1973).
\bibitem{Visser} M. Visser,\textit{ Lorentzian wormholes: From Einstein to Hawking}, (AIP Press, New York, 1995).
\bibitem{phantom} F. S. N. Lobo, Phys. Rev. D {\bf 71}, 084011 (2005).
\bibitem{phantom2}J.A. Gonzalez, F. S. Guzman, N. Montelongo-Garcia, and T. Zannias, Phys. Rev. D {\bf 79}, 064027 (2009).
\bibitem{phantom1}F. S. N. Lobo, F. Parsaei, and N. Riazi, Phys. Rev. D {\bf 87}, 084030 (2013).
 \bibitem{cut} M. Visser, S. Kar and N. Dadhich, Phys. Rev. Lett.  {\bf 90}, 201102 (2003).

%\bibitem{cut1} E. Eiroa and G. Romero, Gen. Rel. Grav. {\bf 36}, 651 (2004).
 \bibitem{cut2} N. M. Garcia, F. S. N. Lobo, and M. Visser, Phys. Rev. D {\bf 86}, 044026 (2012).

\bibitem{cut3}S. D. Forghani, S. H. Mazharimousavi, and M. Halilsoy, Phys. Lett. B {\bf804},  135374 (2020).
 \bibitem{Remo} R. Garattini, and F. S. N. Lobo, Class. Quantum Gravity {\bf24}, 2401 (2007).

\bibitem{variable} F. Parsaei and S. Rastgoo, Phys. Rev. D {\bf 99}, 104037 (2019).

 \bibitem{foad} F. Parsaei and S. Rastgoo,  Eur. Phys. J. C  {\bf 80}, 366 (2020).


%\bibitem{b}M. L. Camera, Phys. Lett. B {\bf 573}, 27 (2003).
  \bibitem{b1} K. A. Bronnikov and Sung-Won Kim, Phys. Rev. D {\bf67}, 064027 (2003).
 \bibitem{b2}F. Parsaei, N. Riazi,  Phys.\ Rev.\ D {\bf 91}, 024015 (2015).
 \bibitem{b3} F. Parsaei, N. Riazi,  Phys.\ Rev.\ D {\bf 102}, 044003 (2020).

\bibitem{Bo}  M. G. Richarte, C. Simeone, Phys. Rev. D {\bf 80}, 104033 (2009).
 \bibitem{Bo1} R. Shaikh, Phys. Rev. D { \bf 98} , 064033 (2018).

\bibitem{quad}F. Duplessis, and D. A. Easson, Phys. Rev. D {\bf 92}, 043516 (2015).
 \bibitem{quad1} H. K. Nguyen , and M. Azreg-Aïnou, Eur. Phys. J. C {\bf 83}, 626 (2023).

\bibitem{Cartan} K. A. Bronnikov and A. M. Galiakhmetov, Grav. Cosmol {\bf21}, 283 (2015).
 \bibitem{Cartan1} M.R. Mehdizadeh and A.H. Ziaie, Phys. Rev. D {\bf 95}, 064049 (2017).
 \bibitem{Cartan2} E. D. Grezia, E. Battista, M. Manfredonia, and G. Miele Eur. Phys. J. Plus. {\bf 132}, 537 (2017).
 \bibitem{Rast} H. Moradpour, N. Sadeghnezhad and H. Hendi, Can. J. Phys. { \bf 95}, 1257 (2017).
 \bibitem{RaR} A. Errehymy, A. Banerjee, O. Donmez, M. Daoud, K. S. Nisar, and A. Abdel-Aty, Gen. Relativ. Gravit {\bf56}, 76 (2024).
 \bibitem{RaR1} E. Battista, S. Capozziello, and A. Errehymy, Eur. Phys. J. C {\bf 84}, 1314 (2024).
 \bibitem{fq} F. Parsaei, S. Rastgoo, and P. K. Sahoo Eur. Phys. J. Plus {\bf 137}, 1083 (2022).
%\bibitem{fq1} M. Calzá, and L. Sebastiani, Eur. Phys. J. C {\bf 83}, 247 (2023).
\bibitem{fq2} S. Rastgoo, and F. Parsaei,  Eur. Phys. J. C  {\bf 84}, 563 (2024).
%\bibitem{fq3} G. Mustafa, Z. Hassan, P. H. R. S.Moraes, and P.K.Sahoo, Phys. Lett. B{\bf 821} 136612 (2021).
\bibitem{fq4} A. Banerjee, A. Pradhan, T. Tangphati, and F. Rahaman, Eur. Phys. J. C {\bf 81}, 1031 (2021).
\bibitem{fq44} Z. Hassan, S. Mandal, and P.K. Sahoo, Fortschr. Phys. {\bf69}, 2100023 (2021).
\bibitem{fq5}S. Kiroriwal, J. Kumar, S.K. Maurya, and S. Ray, Phys. Dark Universe {\bf46}, 101559 (2024).
%\bibitem{fq6} S. Kiroriwal, J. Kumar, S. Kumar Maurya, S. Chaudhary, and A. Aziz, Fortschr. Phys. {\bf72}, 2300197 (2024).
\bibitem{Nojiri}S.~Nojiri, S.~D.~Odintsov, and V.~K.~Oikonomou, Phys. Rept. {\bf692}, 1 (2017).
%\bibitem{fR0} P. Pavlovic, and M. Sossich, Eur. Phys. J. C  {\bf 75}, 117 (2015).
\bibitem{fR11} E. F. Eiroa, and G. F. Aguirre,  Eur. Phys. J. C  {\bf 6}, 132 (2016).
\bibitem{fR22} T. Multam¨aki, I. Vilja, Phys. Rev. D 74, 064022 (2006).
%\bibitem{fR33} F. Parsaei and S. Rastgoo,	arXiv:2110.07278v1.
%\bibitem{fR44}N. Godani,  and G. C. Samanta, Eur. Phys. J. C {\bf 80}, 30 (2020).
\bibitem{fR55}A. S. Agrawal, B. Mishra, F. Tello-Ortiz, and A. Alvarez, Fortschr. Phys. {\bf70}, 2100177 (2022).

\bibitem{inverse}G. Mustafa, Phys. Lett. B {\bf848},  138407 (2024).
\bibitem{Must1} G. Mustafa, M. Ahmad, A. Övgün, M. F. Shamir, and I. Hussain, Fortschr. Phys. {\bf61} ,2100048 (2021).
%\bibitem{Ditta} A. Ditta, I. Hussain, G. Mustafa, A. Errehymy, and M. Daoud, Eur. Phys. J. C {\bf 81}, 880 (2021).
\bibitem{Err}	A. Errehymy , S. Hansraj, S.K. Maurya, C. Hansraj a, M. Daoud,  Phys. Dark Universe {\bf41}, 101258 (2023).
%\bibitem{Chal0} C. C. Chalavadi, N. S. Kavya, and V. Venkatesha, Eur. Phys. J. Plus {\bf138}, 885 (2023).
\bibitem{Riz} M. M. Rizwan, Z. Hassan,P. K. Sahoo, and A. Övgün, Eur. Phys. J. C {\bf 84}, 1132 (2024).
%\bibitem{Chal} C. C. Chalavadi, V. Venkatesha, and A. Malik, Nucl. Phys. B {\bf1006}, 116644 (2024).
\bibitem{Must2} G. Mustafa, A. Errehymy, F. Javed, S.K. Maurya, S. Hansraj, S. Sadiq, J. High Energy Astrophys. {\bf42}, 1 (2024).
\bibitem{foad3}   F. Parsaei, and S. Rastgoo, Ann. Phys. {\bf482}, 170205 (2025).



\bibitem{L1}A. Errehymy , S.K. Maurya, S. Hansraj, M. Mahmoud, K. S. Nisar, and A. Abdel-Aty,  Chinese J. Phys.  {\bf89}, 56 (2024).
\bibitem{L2} T. Naseer, M. Sharif, M. Faiza,  Chin. J. Phys.  {\bf94}, 204 (2025).

%\bibitem{L3} N.S. Kavya, G. Mustafa, V. Venkatesha, P.K. Sahoo, Chinese J. Phys. {\bf87}, 751 (2024).
%\bibitem{L33} S. V. Soni, A. C. Khunt, F. Rahaman, and A.H. Hasmani, arXiv:2503.09222v1.
%\bibitem{L4} T. Naseer, M. Sharif, A. Fatima, and S. Manzoor, Chin. J. Phys. {\bf85}, 350 (2023).
%\bibitem{L5}S. Kiroriwal, J. Kumar, S.K. Maurya, and S. Chaudhary,  Chin. J. Phys.  {\bf89}, 1693 (2024).
%\bibitem{L6}R. Solanki, Z. Hassan, P.K. Sahoo,  Chinese J. Phys.  {\bf85}, 74 (2023).
\bibitem{L7}M. Khatri, Z. Chhakchhuak, and A. Lalchhuangliana, Ann. Phys. {\bf470},  169788 (2024).
%\bibitem{L8}S.K. Maurya, J. Kumar, Sweeti Kiroriwal, and A. Errehymy, Phys. Dark Universe {\bf46},  101564 (2024).
%\bibitem{L9}M. Khatri, and J. Lalvohbika,  Chinese J. Phys.  {\bf89}, 1222 (2024).
\bibitem{L10}G. Mustafa, F. Javed, S. K . Maurya, M. Govender, and A. Saleem, Phys. Dark Universe {\bf45},   101508 (2024).
%\bibitem{L11} L. V. Jaybhaye, M. Tayde, and P. K. Sahoo, Commun. Theor. Phys. {\bf76}, 055402 (2024).
\bibitem{L12} M. Z. Rizwan, Z. Hassan, and P.K. Sahoo Phys. Lett. B {\bf860}, 139152 (2025).
\bibitem{SR2} S. Rastgoo, and F. Parsaei, arXiv:2510.11487v1.

\bibitem{Azizi} T. Azizi, Int. J. Theor. Phys, {\bf 52} 3486 (2013).
\bibitem{Moa} P. H. R. S. Moraes, and P. K. Sahoo, Phys. Rev. D {\bf 96}, 044038 (2017).
\bibitem{Zub} M. Zubair, G. Mustafa, S. Waheed, and G. Abbas,  Eur. Phys. J. C {\bf 77}, 680 (2017).


%\bibitem{Shar} M. Sharif ,and I. Nawazish, Ann. Phys, {\bf 400}, 37 (2019).
%\bibitem{Shw} Shweta, A. K. Mishra, and U. K. Sharma, Int. J. Mod. Phys. D {\bf 35} 2050149 (2020).
%\bibitem{Cha} A. Chanda, S. Dey, and B. C. Paul,   Gen. Relativ. Gravit {\bf 53}, 78 (2021).
%\bibitem{Charge} P. H. R. S. Moraes, W. de Paula, and R. A. C. Correa, Int. J. Mod. Phys. D {\bf 28} 1980098(2019).

%\bibitem{Rosa} J. L. Rosaa, and P. M. Kull, Eur. Phys. J. C {\bf 82}, 1154 (2022).

%\bibitem{Sah} P. Sahoo, S. Mandal, and  P.K. Sahoo, New. Astron. {\bf 80} 101421 (2020).

%\bibitem{squared}S.K. Tripathy , D. Nayak , B. Mishra , D. Behera , and S.K. Sahu, Nucl. Phys. B {\bf1001},  116513 (2024).

%\bibitem{Moa2}P . H. R. S. Moraes, P. K. Sahoo, S. S. Kulkarni, S. Agarwal, Chin. Phys. Lett. {\bf36}, 120401 (2019).
%\bibitem{Zub2}M. Zubair, R. Saleem, Y. Ahmad, and G. Abbas,  Int. J. Geom. Methods. Mod. Phys {\bf16}, 1950046 (2019).
%\bibitem{fr1} E. Elizalde, and M. Khurshudyan, Int. J. Mod. Phys. D {\bf 28}, 1950172 (2019).
\bibitem{fr2} E. Elizalde and M. Khurshudyan, Phys. Rev. D {\bf 98}, 123525 (2018).
%\bibitem{fr3} E. Elizalde and M. Khurshudyan, Phys. Rev. D {\bf99}, 024051 (2019).
%\bibitem{fr4}M. Ilyas, and A. R. Athar, Phys. Scr. {\bf97}, 045003 (2022).
%\bibitem{Yousaf}Z. Yousaf, M. Ilyas, and M. Zaeem-ul-Haq Bhatti, Eur. Phys. J. Plus {\bf132}, 268 (2017).

\bibitem{Sha} U. k. Sharma, abd A. K. Mishra,  Found. Phys {\bf51}, 50 (2021).


\bibitem{Ban}A. Banerjee, M.K. Jasim, and S. G. Ghosh, Ann. Phys, {\bf433}, 16875 (2021).
\bibitem{Sarkar} N. Sarkar, S. Sarkar, A. Bouzenada, A. Dutta, M. Sarkar, and F. Rahaman, Phys. Dark Univ. {\bf44}, 101439 (2024).
\bibitem{SR1} S. Rastgoo, and F. Parsaei, Nucl. Phys. B {\bf1011}, 116797 (2025).
\bibitem{foad4}  F Parsaei and S Rastgoo,  Commun. Theor. Phys. {\bf78}, 025403 (2026).

\bibitem{Ha}T. Harko, F.S.N. Lobo, Eur. Phys. J. C {\bf70}, 373 (2010).
\bibitem{f(T)} Yi-Fu Cai, S. Capozziello, M. D. Laurentis, and E. N. Saridakis, Rept. Prog. Phys. {\bf79}, 106901 (2016).
\bibitem{f(Q)} J. . Jiménez, L. Heisenberg, T. S. Koivisto, and S. PekarPhys. Rev. D {\bf101}, 103507 (2020).
\bibitem{fqt}Y. Xu et al., Eur. Phys. J. C, {\bf79}, 708 (2019).


\bibitem{ap}S. Arora, J. R. L. Santos, and P. K. Sahoo, Phys. Dark Univ. {\bf31}, 100790 (2021).
\bibitem{ap0}M. Tayde, S. Ghosh, and P. K. Sahoo, Chin. Phys. C {\bf47}, 075102 (2023).
\bibitem{ap00}R. Garg, G. P. Singh, A. R. Lalke, and S. Ray, Phys. Lett. A  {\bf525} ,  129937 (2024).
\bibitem{ap1} S. Arora, S.K.J. Pacif, S. Bhattacharjee, and P.K. Sahoo, Phys. Dark Univ. {\bf30},  100664 (2020).
\bibitem{ap2} N. Godani, and G. C. Samanta, Int. J. Geom. Meth. Mod.Phys. {\bf18}, 2150134 (2021).
\bibitem{ap3}V. K. Bhardwaj, and S. Ray, J. H. E. Astrophys. {\bf33},  1 (2022).
\bibitem{ap4}A. Nájera and A. Fajardo, J. C. A. P {\bf03},020 (2022).
\bibitem{ap5}M. Shiravand, S. Fakhry, M. Farhoudi, Phys. Dark Univ. {\bf37},  101106 (2022).
%\bibitem{ap6}G. N. Gadbail, S. Arora, and P.K. Sahoo, Phys. Dark Univ. {\bf37},  101074 (2022).
\bibitem{ap7}Tee-How Loo, M. Koussour, and A. De,  Ann. Phys, {\bf454},  (2023).
\bibitem{ap8} M. Sharif, and I. Ibrar Chin. J. Phys. {\bf89}, 1578 (2024).
\bibitem{ap9}S. Pradhan, S. K. Maurya, P. K. Sahoo, and G. Mustafa, Fortschr. Phys. {\bf72} , 2400092 (2024).
\bibitem{ap10}M. Sharif, I. Ibrar, Phys.Scripta {\bf99},  105034 (2024).
\bibitem{ap11}S. Rathore, S .S. Singh, S. Muhammad, and E. E. Zotos, Eur. Phys. J. C {\bf84}, 1108 (2024).
\bibitem{ap12}A. Errehymy, S.K. Maurya, O. Donmez, Z. Umbetova, J. Rayimbaev, M. M. Khashan, M. R. Eid, Phys. Dark Univ. {\bf50},  102055 (2025).
\bibitem{ap13}J. Ge, L. Ming, Shi-Dong Liang, Hong-Hao Zhang, T. Harko Phys. Rev. D {\bf111}, 124049 (2025).

\bibitem{ap14}A. Pradhan, M. Zeyauddin, A. Dixit, and K. Ghaderi, Universe {\bf11}, 279 (2025).
\bibitem{ap15} S. D. Sadatian, S. M. R. Hosseini, Int. J. Geom. Meth. Mod. Phys. {\bf22}  2450308 (2025).

\bibitem{1} N. S. Kavya, G. Mustafa, and V. Venkatesha, Ann. Phys, {\bf468},  169723  (2024).
\bibitem{11}C. C. Chalavadi., N.S. Kavya, V. Venkatesha,  Eur. Phys. J. Plus {\bf 138}, 885 (2023).
\bibitem{12a}C. C. Chalavadi, V. Venkatesha, N. S. Kavya, and S. V. D. Rashmi, Commun. Theor. Phys. {\bf76}, 025403 (2024).
\bibitem{12} M. Tayde, Z. Hassan, P.K. Sahoo, Chin. J. Phys. {\bf89},  195 (2024).
\bibitem{2} G. G. L. Nashed, W. E. Hanafy, J. C. A. P {\bf09}, 040 (2025).
\bibitem{3} M. Tayde, Z. Hassan, P.K. Sahoo, Nucl. Phys. B {\bf1000}, 116478 (2024).
\bibitem{4} S. D. Sadatian, and S. M. R. Hosseini,  Adv. High Energy Phys. {\bf2024},  3717418 (2024).
\bibitem{5} M. Tayde, Z. Hassan, P.K. Sahoo, and S. Gutt, Chin. Phys. C {\bf46}, 115101 (2022).
\bibitem{6} M. Tayde, J. R. L. Santos, J. N. Araujo, and P. K. Sahoo, Eur. Phys. J. Plus {\bf 138}, 539 (2023).
\bibitem{7} M. Tayde, S. Ghosh, and P.K. Sahoo, Chin. Phys .C  {\bf47}, 075102 (2023).
\bibitem{8} V. Venkatesha, C. C. Chalavadi, N.S. Kavya, and P.K. Sahoo,  New Astron. {\bf105},  102090 (2024).
\bibitem{9} M. Tayde, Z. Hassan, P.K. Sahoo, Phys .Dark Univ. {\bf42},  101288 (2023).
\bibitem{10} A. Sahoo, S. K. Tripathy, B. Mishra, and S. Ray, Eur. Phys. J. C {\bf84}, 325 (2024).


\bibitem{13} M. Zeeshan Gul, M Sharif, S. Shahid, and F. Javed,  Phys. Scr. {\bf99}, 125004 (2024).
\bibitem{14} A. Pradhan, M. Zeyauddin, A. Dixit, and K. Ghaderi, Universe {\bf11}, 279 (2025).
\end{thebibliography}
\end{document}